\documentclass[pdftex,twocolumn,epjc3,preprintnumbers,amsmath,amssymb]{svjour3}

\RequirePackage[T1]{fontenc}

\smartqed  % flush right qed marks, e.g. at end of proof

\RequirePackage{graphicx}
\RequirePackage{mathptmx}   % use Times fonts if available on your TeX system
\RequirePackage{flushend}
\RequirePackage[numbers,sort&compress]{natbib}
\RequirePackage[colorlinks,citecolor=blue,urlcolor=blue,linkcolor=blue]{hyperref}

\journalname{Eur. Phys. J. C}
\usepackage{amsmath}
\usepackage{amssymb}
\usepackage[normalem]{ulem}
\usepackage{graphicx}
\usepackage{bm}
\usepackage{color}
\usepackage{multicol}
\newcommand{\n}{\nonumber}

\begin{document}
\onecolumn
\title{Explorations of two empirical formulae for fermion masses}
\author{Guan-Hua Gao\thanksref{addr1}
        \and
        Nan Li\thanksref{e3,addr1}}
\thankstext{e3}{Corresponding author. E-mail: linan@mail.neu.edu.cn}
\institute{Department of Physics, College of Sciences, Northeastern University, Shenyang 110819, China\label{addr1}}
\date{Received: date / Accepted: date}
\maketitle

\begin{abstract}
Two empirical formulae for the lepton and quark masses (i.e. Kartavtsev's extended Koide formulae), $K_l=(\sum_l m_l)/(\sum_l\sqrt{m_l})^2=2/3$ and $K_q=(\sum_q m_q)/(\sum_q\sqrt{m_q})^2=2/3$, are explored in this paper. For the lepton sector, we show that $K_l=2/3$, only if the uncertainty of the tauon mass is relaxed to about $2\sigma$ confidence level, and the neutrino masses can consequently be extracted with the current experimental data. For the quark sector, the extended Koide formula should only be applied to the running quark masses, and $K_q$ is found to be rather insensitive to the renormalization effects in a large range of energy scales from GeV to $10^{12}$ GeV. We find that $K_q$ is always slightly larger than $2/3$, but the discrepancy is merely about $5\%$.

\end{abstract}

\section{Introduction} \label{sec:intro}

Despite the glorious successes of the standard model of particle physics, the generations of fermion masses remain one of the most fundamental but unsolved problems therein. These masses are treated as free parameters in the standard model, which seem to be rather dispersed and unrelated and can only be determined experimentally. Therefore, it is reasonable to firstly seek some phenomenological relations of these masses in order to reduce the number of free parameters, and this will significantly help us for the future model buildings in and beyond the standard model.

Among the existing phenomenological mass relations, an empirical formula suggested by Koide three decades ago \cite{Koide}, is one of the most accurate,
\begin{align}
k_l=\frac{m_e+m_\mu+m_\tau}{(\sqrt{m_e}+\sqrt{m_\mu}+\sqrt{m_\tau})^2}=\frac23, \n
\end{align}
where $m_e$, $m_\mu$, and $m_\tau$ are the pole masses of three charged leptons: electron, muon, and tauon, respectively. From the current experimental data [best-fit ($\pm 1\sigma$)] \cite{PDG}:
\begin{align}
m_e=(0.510998928 \pm 0.000000011)~{\rm MeV}, \quad m_\mu=(105.6583715 \pm 0.0000035)~{\rm MeV}, \quad m_\tau=(1776.82 \pm 0.16)~{\rm MeV}, \n
\end{align}
it is straightforward to see the amazing precision of this simple formula,
\begin{align}
k_l=\frac23\times\left[1\pm\mathcal{O}\left(10^{-5}\right)\right]. \n
\end{align}
Besides, it is very interesting to see that $k_l$ lies exactly in the middle of the two extremes of $1/3$ (exact democracy in lepton mass spectrum) and $1$ (extreme hierarchy) \cite{Gerard:2005ad}. Furthermore, although the lepton masses are hierarchical, the exclusion of the smallest mass $m_e$ still makes $k_l$ notably deviate from $2/3$.

This remarkable precision has aroused in both theorists \cite{Koide:1989jq,Koide:1995xk,Gerard:2005ad,Xing:2006vk,Ma:2006ht,Sumino,Uekusa:2009pj,z} and phenomenologists \cite{Foot,Li:2005,Rodejohann:2011jj,K} a longtime interest in the Koide formula, but unfortunately the underlying physics remains incomplete. The previous studies can be classified into three categories: (1) to explore the possible physical origin of the Koide formula (e.g. in a supersymmetric model \cite{Ma:2006ht} or in an effective field theory \cite{Sumino}); (2) to generalize the Koide formula from charged leptons to neutrinos and quarks (see Refs. \cite{Li:2005} and \cite{Rodejohann:2011jj} for example); (3) to examine the energy scale dependence of the Koide formula \cite{Xing:2006vk} (i.e. to check the stability of the Koide formula against radiative corrections, with the pole masses replaced by the running masses of fermions). In the present paper, we focus on the latter two aspects for two extended Koide formulae suggested by Kartavtsev in Ref. \cite{K}. Before doing so, we briefly review the relevant works on the extensions of the Koide formula from charged leptons to neutrinos and quarks.

In Ref. \cite{Li:2005}, the Koide formula was extended to the sectors of neutrinos, up-type, and down-type quarks,
\begin{align}
k_\nu=\frac{m_1+m_2+m_3}{(\sqrt{m_1}+\sqrt{m_2}+\sqrt{m_3})^2}, \quad
k_{\rm up}=\frac{m_u+m_c+m_t}{(\sqrt{m_u}+\sqrt{m_c}+\sqrt{m_t})^2}, \quad
k_{\rm down}=\frac{m_d+m_s+m_b}{(\sqrt{m_d}+\sqrt{m_s}+\sqrt{m_b})^2}, \label{li}
\end{align}
where $m_1$, $m_2$, $m_3$ are the masses of three neutrinos, and $m_u$, $\cdots$, $m_b$ are the masses of six quarks. It was found that these naive extensions of the Koide formula failed to be valid, as $k_{\nu}=0.33\sim 0.57$, $k_{\rm up}=0.73\sim 0.93$, and $k_{\rm down}=0.60\sim 0.80$. In Ref. \cite{Rodejohann:2011jj}, Rodejohann and Zhang generalized the Koide formula to the quark sector in another way, not according to their charges or isospins, but to their masses. For the light ($u$, $d$, $s$) and heavy ($c$, $b$, $t$) quarks, they separately introduced
\begin{align}
k_{\rm light}=\frac{m_u+m_d+m_s}{(\sqrt{m_u}+\sqrt{m_d}+\sqrt{m_s})^2}, \quad
k_{\rm heavy}=\frac{m_c+m_b+m_t}{(\sqrt{m_c}+\sqrt{m_b}+\sqrt{m_t})^2}, \label{zhang}
\end{align}
and found a better accordance with the Koide formula, $k_{\rm light}=0.49\sim0.65$ and $k_{\rm heavy}=0.66\sim0.68$.

Recently, Kartavtsev proposed a new extension \cite{K}, including the neutrino masses in the original Koide formula,
\begin{align}
K_l=\frac{\sum_l m_l}{(\sum_l\sqrt{m_l})^2} =\frac{m_1+m_2+m_3+m_e+m_\mu+m_\tau}{(\sqrt{m_1}+\sqrt{m_2}+\sqrt{m_3}+\sqrt{m_e}+\sqrt{m_\mu}+\sqrt{m_\tau})^2}, \label{l}
\end{align}
where the sum is democratically over all the six leptons. Since the neutrino masses are tiny compared to those of charged leptons, it is quite possible that the exact ratio $2/3$ could be attained when they are taken into account. Another motivation for this extension is from the quark sector,
\begin{align}
K_q=\frac{\sum_q m_q}{(\sum_q\sqrt{m_q})^2}
=\frac{m_u+m_d+m_s+m_c+m_b+m_t}{(\sqrt{m_u}+\sqrt{m_d}+\sqrt{m_s}+\sqrt{m_c}+\sqrt{m_b}+\sqrt{m_t})^2}. \label{kq}
\end{align}
It will be shown in Sect. \ref{sec:qua} that $K_q$ is also very close to $2/3$ and is almost stable in a large range of energy scales, although the mass of each quark varies with the energy scale obviously.

These preliminary successes enlighten us to investigate Kartavtsev's extensions of the Koide formula in detail, since the naive estimates in Ref. \cite{K} were still very simple. We should firstly explore phenomenologically the validity of Eqs. (\ref{l}) and (\ref{kq}) for both leptons and quarks, and these explorations will be greatly helpful for the relevant model building for the theoretical explanation of the extended Koide formulae. This is the purpose of our present paper.

This paper is organized as follows. First, in Sect. \ref{sec:lep}, we show that the extended Koide formula for leptons is invalid under the constraints from current experimental data. However, if the uncertainty of the most inaccurate tauon mass is relaxed to about $2\sigma$ confidence level, the extended Koide formula could be satisfied, and the neutrino masses can thus be extracted. Next, in Sect. \ref{sec:qua}, we first clarify some different definitions of quark masses and then examine the energy scale dependence of the extended Koide formula for the running quark masses. We find that the extended Koide formula holds fairly well in a large range of energy scales. Some relevant conclusions and discussions are shown in Sect. \ref{sec:dis}.

\section{Extended Koide formula for leptons} \label{sec:lep}

In this section, we examine the extended Koide formula for leptons. However, this can only be performed with the masses of all the six leptons known. Unfortunately, the lack of the absolute masses of neutrinos makes this examination impossible. But on the other hand, thanks to the more and more precise experiments of neutrino oscillations, we have already had two firm constraints of the mass-squared differences of the neutrino mass eigenstates. Therefore, if we assume the validity of the extended Koide formula for leptons (i.e $K_l=2/3$ exactly) and regard it as the third constraint, the three neutrino masses may thus be extracted.

Based on 3-flavor neutrino mixing, a global analysis \cite{PDG} of the current experimental data from the oscillations of the solar, atmospheric, reactor, and accelerator neutrinos indicates [best-fit ($\pm 1\sigma$)]:
\begin{align}
m_2^2-m_1^2&=(7.53\pm0.18)\times10^{-5}~{\rm eV}^2, \n\\
m_3^2-m_2^2&=(2.44\pm0.06)\times10^{-3}~{\rm eV}^2 \quad ({\rm normal~mass~hierarchy}), \n\\
m_2^2-m_3^2&=(2.52\pm0.07)\times10^{-3}~{\rm eV}^2 \quad ({\rm inverted~mass~hierarchy}). \n
\end{align}
By normal and inverted mass hierarchies, we mean that neutrinos have the mass spectra $m_1<m_2<m_3$ and $m_3<m_1<m_2$, respectively.

Suppose the extended Koide formula is exact for leptons,
\begin{align}
K_l=\frac{m_1+m_2+m_3+m_e+m_\mu+m_\tau}{(\sqrt{m_1}+\sqrt{m_2}+\sqrt{m_3}+\sqrt{m_e}+\sqrt{m_\mu}+\sqrt{m_\tau})^2}=\frac23. \label{kl}
\end{align}
Solving Eq. (\ref{kl}) with the two constraints of the mass-squared differences, we may first determine one of the three neutrino masses (e.g. $m_2$) and then obtain the other two ($m_1$ and $m_3$). But the direct solving of Eq. (\ref{kl}) is rather inconvenient, so we convert this problem to search the local minimum of $|K_l-2/3|$, and this is a typical numerical nonlinear optimization problem. In the following, we synthetically utilize the Nelder--Mead method (with the shortest runtime) and the random search method (with the highest accuracy) to obtain the neutrino masses with the smallest computational cost.

We start from the normal mass hierarchy. In this circumstance, what we face now is to search the minimum of $|K_l-2/3|$ as an objective function of six variables ($m_1$, $m_2$, $m_3$, $m_e$, $m_\mu$, $m_\tau$) under five constraints:
\begin{align}
(7.53-0.18)\times10^{-5}~{\rm eV}^2&< m_2^2-m_1^2< (7.53+0.18)\times10^{-5}~{\rm eV}^2, \label{12}\\
(2.44-0.06)\times10^{-3}~{\rm eV}^2&< m_3^2-m_2^2< (2.44+0.06)\times10^{-3}~{\rm eV}^2, \label{23}\\
(0.510998928 - 0.000000011)~{\rm MeV}&< m_e<(0.510998928 + 0.000000011)~{\rm MeV}, \label{e}\\
(105.6583715 - 0.0000035)~{\rm MeV}&< m_\mu<(105.6583715 + 0.0000035)~{\rm MeV}, \label{mu}\\
(1776.82 - 0.16)~{\rm MeV}&< m_\tau<(1776.82 + 0.16)~{\rm MeV}. \label{tau}
\end{align}
For searching the minimum of $|K_l-2/3|$, we set the precision of operation to $30$ significant figures. Meanwhile, due to the limitations of the Nelder--Mead method, a computing result has a great possibility to oscillate in the vicinity of a convergence point, so we set the maximal number of iterations to $2000$ for a single operation. If the iterations are not successful, the point will be discarded, and the results from this point will be excluded by the checking program. From the operating and checking for $500$ random initial points, we eventually obtain $49$ successful convergence points, and the first $10$ minima of $|K_l-2/3|$ and the corresponding neutrino masses are sorted in Tab. \ref{leptons}.
\begin{table}[h]
\begin{tabular}{c|ccc|ccc}
  \hline
$|K_l-2/3|$ & $m_1$(eV)& $m_2$ (eV) & $m_3$ (eV)& $m_e$ (MeV)& $m_\mu$ (MeV) & $m_\tau$ (MeV) \\
\hline
$7.28894\times 10^{-6}$ & $4.89844\times 10^{-8}$ & $8.58877\times 10^{-3}$ & $4.95356\times 10^{-2}$ & 0.510999 & 105.658 & 1776.98 \\
$7.29051\times 10^{-6}$ & $1.85238\times 10^{-8}$ & $8.58196\times 10^{-3}$ & $4.95343\times 10^{-2}$ & 0.510999 & 105.658 & 1776.98 \\
$7.29142\times 10^{-6}$ & $9.25236\times 10^{-8}$ & $8.59159\times 10^{-3}$ & $4.95360\times 10^{-2}$ & 0.510999 & 105.658 & 1776.98 \\
$7.30195\times 10^{-6}$ & $2.81761\times 10^{-7}$ & $8.62488\times 10^{-3}$ & $4.95420\times 10^{-2}$ & 0.510999 & 105.658 & 1776.98 \\
$7.31602\times 10^{-6}$ & $6.75381\times 10^{-8}$ & $8.60216\times 10^{-3}$ & $4.99676\times 10^{-2}$ & 0.510999 & 105.658 & 1776.98 \\
$7.32170\times 10^{-6}$ & $3.37241\times 10^{-10}$& $8.57321\times 10^{-3}$ & $5.02470\times 10^{-2}$ & 0.510999 & 105.658 & 1776.98 \\
$7.35571\times 10^{-6}$ & $6.34196\times 10^{-6}$ & $8.58244\times 10^{-3}$ & $4.97131\times 10^{-2}$ & 0.510999 & 105.658 & 1776.98 \\
$8.32900\times 10^{-6}$ & $1.67259\times 10^{-3}$ & $8.73485\times 10^{-3}$ & $4.95610\times 10^{-2}$ & 0.510999 & 105.658 & 1776.98 \\
$1.00832\times 10^{-5}$ & $6.54993\times 10^{-3}$ & $1.07890\times 10^{-2}$ & $4.99830\times 10^{-2}$ & 0.510999 & 105.658 & 1776.97 \\
$1.71168\times 10^{-5}$ & $4.87567\times 10^{-2}$ & $4.95407\times 10^{-2}$ & $6.95290\times 10^{-2}$ & 0.510999 & 105.658 & 1776.98 \\
  \hline
\end{tabular}
\caption{The first 10 minima of $|K_l-2/3|$ and the corresponding neutrino masses from the extended Koide formula in the normal mass hierarchy. The values of $|K_l-2/3|$ are greatly larger than the precision of operation, meaning that the extended Koide formula for leptons is unsuccessful under the constraints from current experimental data.} \label{leptons}
\end{table}

From Tab. \ref{leptons}, we clearly observe that although the values of $|K_l-2/3|$ are already very small $\sim\mathcal{O}(10^{-6})$, they are still greatly beyond the computing accuracy that we adopt. Since we have set the precision of operation to 30 significant figures, $|K_l-2/3|$ should be $\sim\mathcal{O}(10^{-30})$, if the masses $m_1$, $m_2$, $m_3$ make $|K_l-2/3|$ converge to zero absolutely. As a result, we conclude that there is no solution for neutrino masses from Eq. (\ref{kl}) under the current experimental constraints in Eqs. (\ref{12})--(\ref{tau}). In other words, Kartavtsev's extended Koide formula for leptons in Eq. (\ref{kl}) is unsuccessful. This situation is the same for the inverted mass hierarchy.

\vskip .3cm

\begin{figure}[h]
\begin{center}
\includegraphics[width=0.45\linewidth,angle=0]{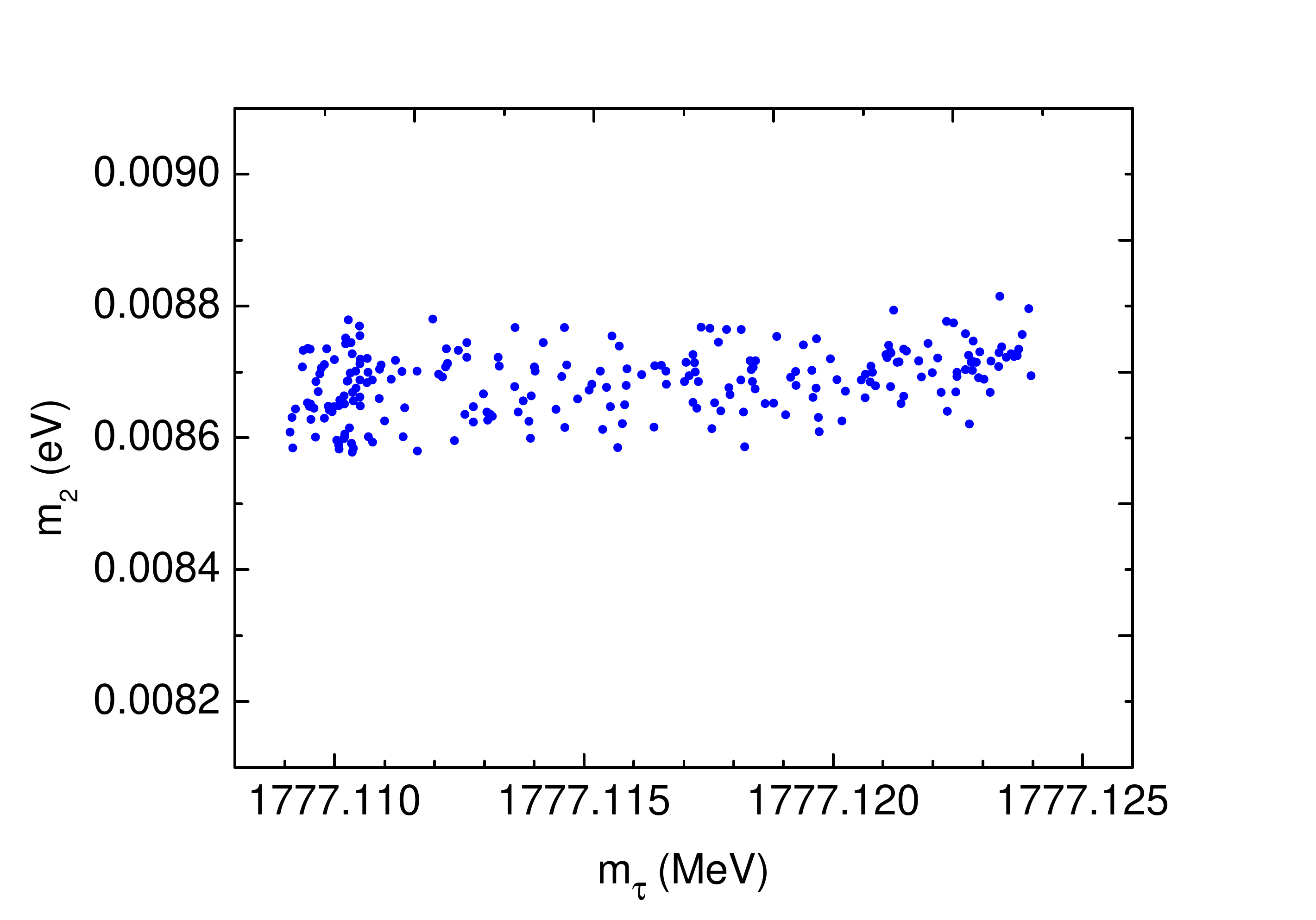} \qquad \includegraphics[width=0.45\linewidth,angle=0]{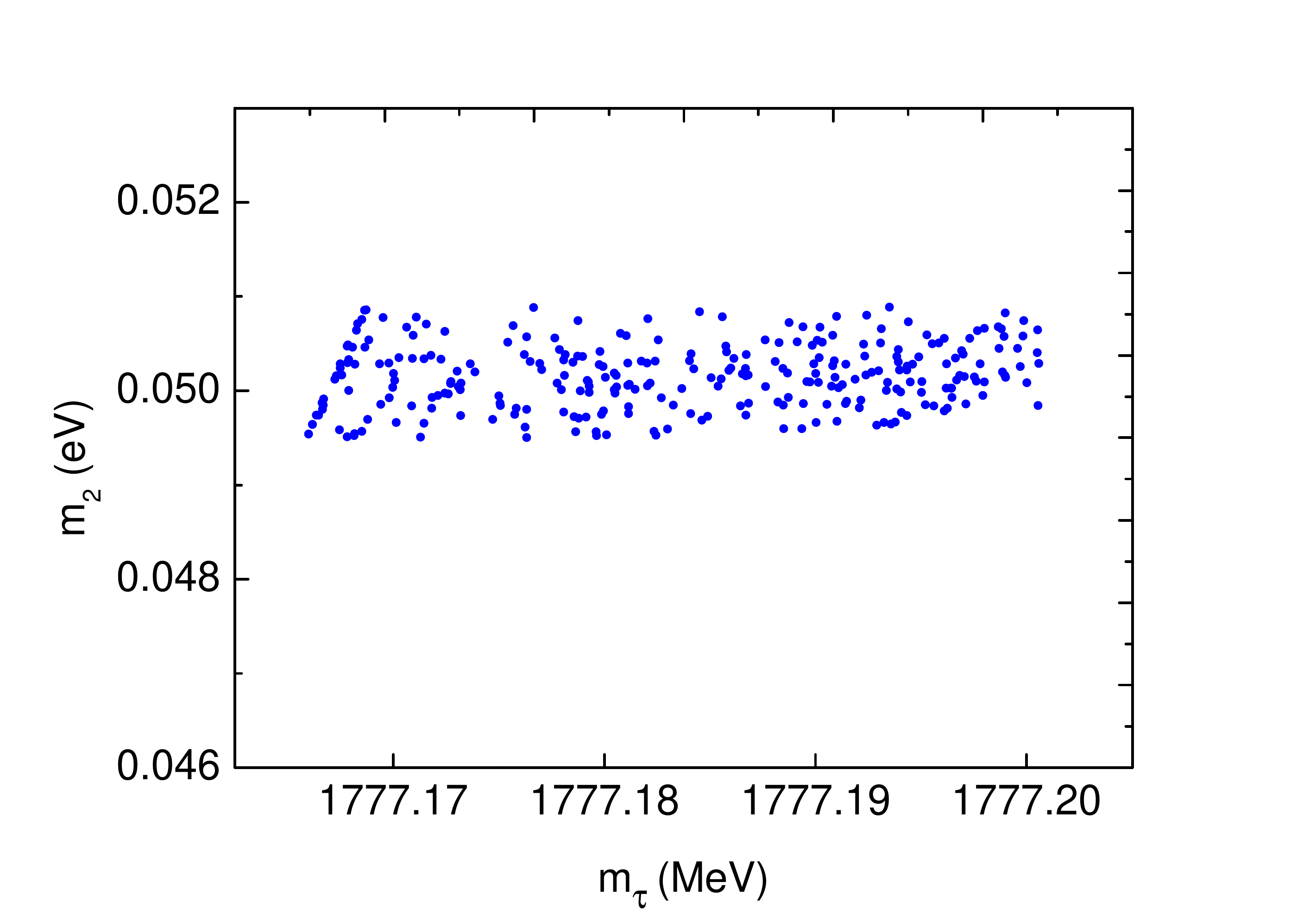}
\end{center}
\caption{The neutrino mass $m_2$ and its dependence on the tauon mass $m_\tau$. The left and right panels are for the normal and inverted mass hierarchies, respectively. Each point in the figures is a convergence point for $K_l=2/3$. There is a lower bound 1777.11 MeV for $m_\tau$ in the normal mass hierarchy and 1777.17 MeV in the inverted case. The neutrino mass $m_2$ is not very sensitive to $m_\tau$ in a relative large range, with the central value being $8.68\times 10^{-3}$ eV in the normal mass hierarchy, and $5.02\times 10^{-2}$ eV in the inverted case.} \label{fig}
\end{figure}

Although the extended Koide formula is invalid for leptons with the current experimental data, the deviations of $K_l$ from $2/3$ are still considerably small. This smallness naturally leads us to investigate weather $K_l=2/3$ exactly, if the uncertainties of lepton masses are slightly larger than the current data. Whereas, we do not need to relax the uncertainty of each lepton mass to test this possibility, because the uncertainty of $K_l$ is substantially attributed to the uncertainty of the tauon mass $m_\tau$, since the uncertainties of $m_1$, $m_2$, $m_3$, $m_e$, and $m_\mu$ are much smaller. In fact, the measurement of $m_\tau$ is much more inaccurate than that shown in Eq. (\ref{tau}): $m_\tau=(1776.82 \pm 0.16)$ MeV, which results from the weighted average of various experiments \cite{PDG}. Actually, even the central value of $m_\tau$ varies from 1775 to 1783 MeV, with the error bar of a few MeV \cite{PDG}. Therefore, in the following, we keep the constraints in Eqs. (\ref{12})--(\ref{mu}) and also the central value of $m_\tau$ as 1776.82 MeV, but only relax its uncertainty, in order to examine the validity of the extended Koide formula. We choose the neutrino mass $m_2$ and study its dependence on $m_\tau$, as $m_2$ appears in both the constraints in Eqs. (\ref{12}) and (\ref{23}). Our results are shown in Fig. \ref{fig}.

From Fig. \ref{fig}, we obviously find that there is a lower bound for the tauon mass $m_\tau$, such that $K_l=2/3$ exactly. For the normal mass hierarchy, from the left panel in Fig. \ref{fig}, we observe
\begin{align}
m_\tau>1777.11~{\rm MeV}, \quad {\rm i.e.} \quad m_\tau>(1776.82+1.81\sigma)~{\rm MeV}, \n
\end{align}
meaning that the extended Koide formula for leptons is satisfied, only if the uncertainty of $m_\tau$ is relaxed to about $2\sigma$ confidence level. Furthermore, the neutrino mass $m_2$ is not very sensitive to $m_\tau$ in a relative large range, and we thus estimate
\begin{align}
8.58\times 10^{-3}~{\rm eV}<m_2< 8.78\times 10^{-3}~{\rm eV}. \n
\end{align}
Therefore, combining Eqs. (\ref{12}) and (\ref{23}), we have the masses of the other two neutrinos as
\begin{align}
0~{\rm eV} <m_1<6.64\times 10^{-4}~{\rm eV}, \quad 4.95\times 10^{-2}~{\rm eV}< m_3< 5.07\times 10^{-2}~{\rm eV}. \n
\end{align}
The corresponding central values for these three neutrino masses read
\begin{align}
m_1=2.06\times 10^{-4}~{\rm eV}, \quad m_2=8.68\times 10^{-3}~{\rm eV}, \quad m_3=5.02\times 10^{-2}~{\rm eV}. \n
\end{align}
Hence, we find a relatively mild hierarchy for the neutrino masses in this case.

Similarly, for the inverted mass hierarchy, from the right panel in Fig. \ref{fig}, we have
\begin{align}
m_\tau>1777.17~{\rm MeV}, \quad {\rm i.e.} \quad m_\tau>(1776.82+2.16\sigma)~{\rm MeV}, \n
\end{align}
and
\begin{align}
0~{\rm eV}< m_3< 3.93\times 10^{-3}~{\rm eV}, \quad
4.87\times 10^{-2}~{\rm eV} <m_1<5.02\times 10^{-2}~{\rm eV}, \quad
4.95\times 10^{-2}~{\rm eV} <m_2<5.09\times 10^{-2}~{\rm eV}, \n
\end{align}
with the central values being
\begin{align}
m_3=2.00\times 10^{-4}~{\rm eV}, \quad m_1=4.94\times 10^{-2}~{\rm eV}, \quad m_2=5.02\times 10^{-2}~{\rm eV}. \n
\end{align}
In this case, we find the neutrino mass scheme as $m_3\ll m_1\approx m_2$.

Last, we should stress that we only focus on the pole masses of leptons in this section, but not their running masses. Since the lepton mass ratios are rather insensitive to radiative corrections \cite{Xing:2006vk}, this is not a severe problem.

\section{Extended Koide formula for quarks} \label{sec:qua}

In this section, we move on to explore the extended Koide formula for quarks (i.e. if $K_q=2/3$ or not). However, the situation in the quark sector is much more complicated, or even ambiguous, than that of lepton. This complication or ambiguity comes from what masses we mean for quarks. For instance, in Ref. \cite{PDG}, the six quark masses are recommended as
\begin{align}
m_u&=2.3^{+0.7}_{-0.5}~{\rm MeV}, \quad m_d=4.8^{+0.5}_{-0.3}~{\rm MeV}, \quad m_s=95\pm 5~{\rm MeV}, \n\\
m_c&=1.275\pm 0.025~{\rm GeV}, \quad m_b=4.18\pm 0.03~{\rm GeV}, \quad m_t=173.21\pm 0.51\pm 0.71~{\rm GeV}. \label{qm}
\end{align}
From these data, it is easy to find $K_q\approx 0.64$. But this trivial result actually makes no sense, as the masses of the light quarks ($u$, $d$, $s$) are estimated as the current quark masses, the masses of relatively heavy quarks ($c$, $b$) mean the running quark masses, and the mass of the heaviest $t$ quark is measured as its pole mass. Therefore, it is meaningless to calculate $K_q$ as a combination of all these six masses without distinction.

Different from leptons, quarks are confined inside hadrons and cannot be observed as physical particles in experiments, so their masses cannot be measured directly. Therefore, concerning quark masses, we should first make clear their quantitative definitions and meanings. For example, the masses of light quarks in chiral perturbation theory always mean the current quark masses. While, in a particular non-relativistic hadron model, we mean the quark masses by the constituent quark masses. Moreover, the quark masses computed directly from lattice quantum chromodynamics (QCD) are the bare quark masses. Whereas, in the Koide formula, the charged lepton masses are the pole (physical) masses, which correspond to the positions of divergence in their propagators in the on-shell renormalization scheme. However, the pole masses of quarks can only be defined in perturbation theory and are not reliable at low energies because of the non-perturbative infrared effects in QCD. Hence, the pole masses of quarks are not well-defined, so the extension of the Koide formula for quarks should only refer to their running masses.

The mass parameters in the QCD Lagrangian depend not only on the renormalization scheme adopted to define the theory, but also on the energy scale at which an observation occurs. Therefore, the values of quark masses may alter significantly in different renormalization schemes and at different energy scales. Whereas, one can convert these values between different schemes in perturbation theory and run these values to the demanded energy scales by the renormalization group equations. At high energies, where non-perturbative QCD effects become small, the calculations of quark masses are most commonly performed in the dimensional regularization scheme with the modified minimal subtraction ($\overline{{\rm MS}}$), to obtain the running (or renormalized) quark masses $m_q(\mu)$ at a given energy scale $\mu$. As a result, we should trace the energy scale dependence of the extended Koide formula for the quark sector and clarify the mass parameters in Eq. (\ref{kq}) as the running quark masses,
\begin{align}
K_q(\mu)=\frac{m_u(\mu)+m_d(\mu)+m_s(\mu)+m_c(\mu)+m_b(\mu)+m_t(\mu)}
{\left[\sqrt{m_u(\mu)}+\sqrt{m_d(\mu)}+\sqrt{m_s(\mu)}+\sqrt{m_c(\mu)}+\sqrt{m_b(\mu)}+\sqrt{m_t(\mu)}\right]^2}. \n
\end{align}
The running quark masses were systematically calculated in Refs. \cite{Koideq} and \cite{Xing}, and were especially recalculated with a much higher precision in Ref. \cite{Xinga} after the discovery of the Higgs boson. Hence, in this paper, we follow the data in Ref. \cite{Xinga}.

In Tab. \ref{quarks}, we list the running quark masses taken from Tab. I of Ref. \cite{Xinga}. These masses were calculated in the standard model at a number of typical energy scales: for example, $m_c$ evaluated at the scale equal to its mass, 2 GeV where light quark masses are often quoted in the $\overline{{\rm MS}}$ scheme, the Higgs mass $m_H\approx 125$ GeV, until the cutoff scale $\Lambda_{\rm VS}\approx 4\times 10^{12}$ GeV, where the vacuum stability in the standard model is lost due to a relatively small Higgs mass. We clearly see that the running quark masses monotonously decrease at large energy scales. The running parameter $K_q(\mu)$ in the extended Koide formula is also listed in the last column in Tab. \ref{quarks}.
\begin{table}[h]
\begin{tabular}{c|ccc|ccc|c}
  \hline
$\mu$              & $m_u(\mu)$ (MeV) & $m_d(\mu)$ (MeV) & $m_s(\mu)$ (MeV) & $m_c(\mu)$ (GeV) & $m_b(\mu)$ (GeV) & $m_t(\mu)$ (GeV) & $K_q(\mu)$ \\
  \hline
$m_c(m_c)$         & $2.79^{+0.83}_{-0.82}$ & $5.69^{0.96+}_{-0.95}$ & $116^{+36}_{-24}$ & $1.29^{+0.05}_{-0.11}$    & $5.95^{+0.37}_{-0.15}$  & $385.7^{+8.1}_{-7.8}$ & $0.701^{+0.010}_{-0.011}$ \\
2 GeV              & $2.4^{+0.7}_{-0.7}$    & $4.9^{+0.8}_{-0.8}$    & $100^{+30}_{-20}$ & $1.11^{+0.07}_{-0.14}$    & $5.06^{+0.29}_{-0.11}$  & $322.2^{+5.0}_{-4.9}$ & $0.698^{+0.010}_{-0.011}$ \\
$m_b(m_b)$         & $2.02^{+0.60}_{-0.60}$ & $4.12^{+0.69}_{-0.68}$ & $84^{+26}_{-17}$  & $0.934^{+0.058}_{-0.120}$ & $4.19^{+0.18}_{-0.16}$  & $261.8^{+3.0}_{-2.9}$ & $0.696^{+0.011}_{-0.009}$ \\
$m_W$              & $1.39^{+0.42}_{-0.41}$ & $2.85^{+0.49}_{-0.48}$ & $58^{+18}_{-12}$  & $0.645^{+0.043}_{-0.085}$ & $2.90^{+0.16}_{-0.06}$  & $174.2^{+1.2}_{-1.2}$ & $0.691^{+0.010}_{-0.010}$ \\
$m_Z$              & $1.38^{+0.42}_{-0.41}$ & $2.82^{+0.48}_{-0.48}$ & $57^{+18}_{-12}$  & $0.638^{+0.043}_{-0.084}$ & $2.86^{+0.16}_{-0.06}$  & $172.1^{+1.2}_{-1.2}$ & $0.692^{+0.010}_{-0.010}$ \\
$m_H$              & $1.34^{+0.40}_{-0.40}$ & $2.74^{+0.47}_{-0.47}$ & $56^{+17}_{-12}$  & $0.621^{+0.041}_{-0.082}$ & $2.79^{+0.15}_{-0.06}$  & $167.0^{+1.2}_{-1.2}$ & $0.691^{+0.010}_{-0.010}$ \\
$m_t(m_t)$         & $1.31^{+0.40}_{-0.39}$ & $2.68^{+0.46}_{-0.46}$ & $55^{+17}_{-11}$  & $0.608^{+0.041}_{-0.080}$ & $2.73^{+0.15}_{-0.06}$  & $163.3^{+1.1}_{-1.}$ & $0.691^{+0.010}_{-0.010}$ \\
1 TeV              & $1.17^{+0.35}_{-0.35}$ & $2.40^{+0.42}_{-0.41}$ & $49^{+15}_{-10}$  & $0.543^{+0.037}_{-0.072}$ & $2.41^{+0.14}_{-0.05}$  & $148.1^{+1.3}_{-1.3}$ & $0.693^{+0.010}_{-0.010}$ \\
$\Lambda_{\rm VS}$ & $0.61^{+0.19}_{-0.18}$ & $1.27^{+0.22}_{-0.22}$ & $26^{+8}_{-5}$    & $0.281^{+0.02}_{-0.04}$   & $1.16^{+0.07}_{-0.02}$  & $82.6^{+1.4}_{-1.4}$  & $0.705^{+0.011}_{-0.011}$ \\
  \hline
\end{tabular}
\caption{The running quark masses and the running parameter $K_q(\mu)$ in the extended Koide formula at some typical energy scales. The mass of the Higgs boson is taken as 125 GeV, and the cutoff scale for vacuum stability is $4\times 10^{12}$ GeV (data from Tab. I of Ref. \cite{Xinga}). The running quark masses are found to decrease monotonically, but $K_q(\mu)$ is almost stable in a sizable range of energy scales from GeV to $10^{12}$ GeV. Moreover, $K_q(\mu)>2/3$ at all energy scales, but the deviations are only about $5\%$.} \label{quarks}
\end{table}

\begin{figure}[h]
\begin{center}
\includegraphics[width=0.5\linewidth,angle=0]{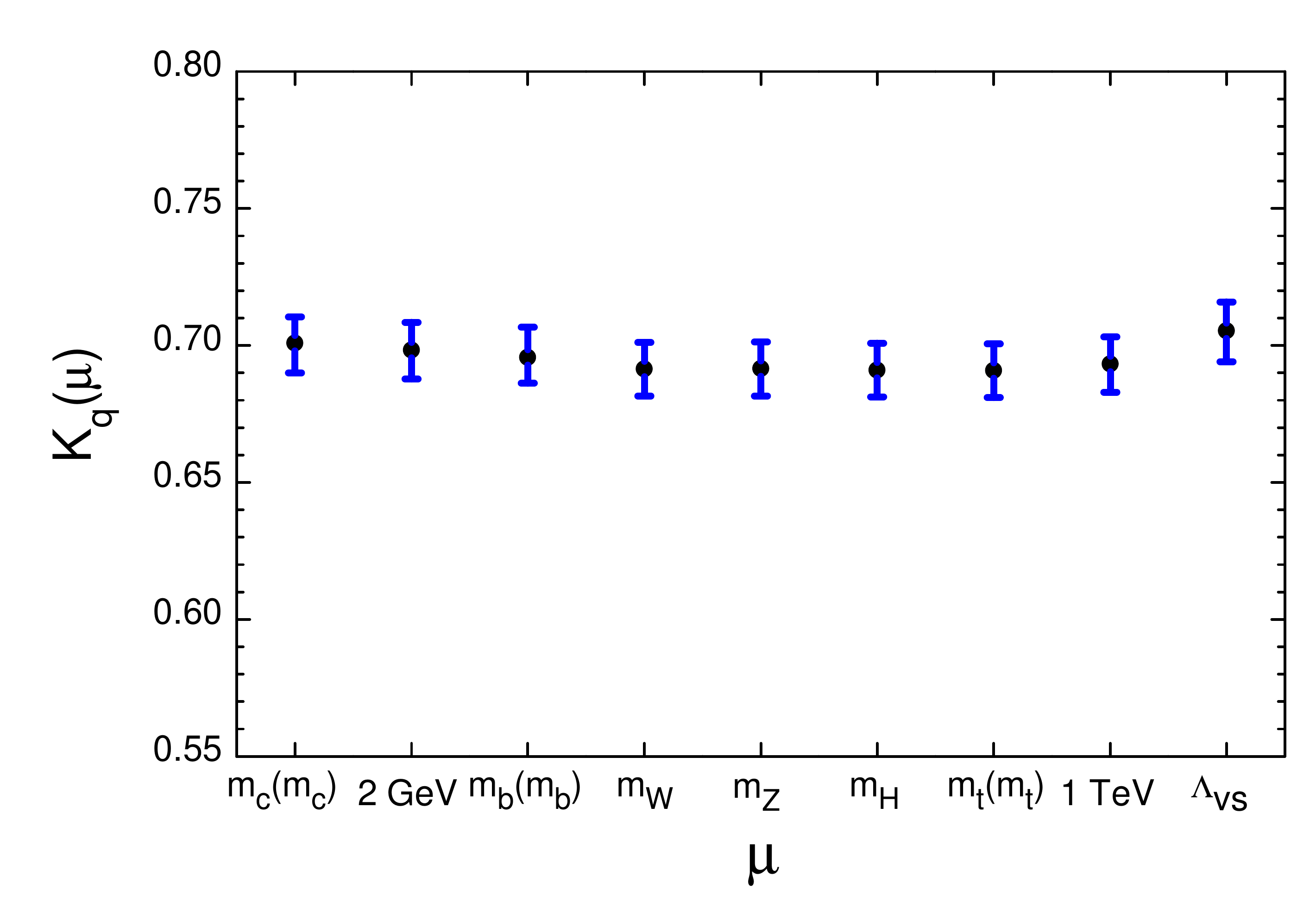}
\end{center}
\caption{The running parameter $K_q(\mu)$ in the extended Koide formula as the function of energy scales. Due to the huge differences between these energy scales, we arrange them equidistantly on the $\mu$-axis. $K_q(\mu)$ is found to be almost stable in a huge range of energy scales from GeV to $10^{12}$ GeV.} \label{k}
\end{figure}

We find from Tab. \ref{quarks} that in a sizable range of energy scales from GeV to $10^{12}$ GeV, $K_q(\mu)$ is rather insensitive to the running effects of quark masses. This insensitivity should be substantially attributed to the large mass hierarchy in the quark sector. Moreover, $K_q(\mu)$ is always slightly larger than $2/3$ at all energy scales, indicating the impossibility to extend the Koide formula to the running quark masses. But the discrepancies between $K_q(\mu)$ and $2/3$ are only about $5\%$, much smaller than the various attempts in Eqs. (\ref{li}) or (\ref{zhang}). In this sense, Kartavtsev's extended Koide formula for quarks is a rather good choice. In addition, the uncertainties of $K_q(\mu)$ at different energy scales are almost the same, but this is not unexpected, as the uncertainty of $K_q(\mu)$ is mainly from the uncertainty of the heaviest $t$ quark mass. Besides, we do not find that $K_q(\mu)$ crosses $2/3$ at some particular energy scale $2~{\rm GeV}<\mu<m_Z$, as claimed by Kartavtsev in Ref. \cite{K}, since this cross was naively estimated from the data in Eq. (\ref{qm}). But as we have explained above, these values of quark masses in Eq. (\ref{qm}) have different definitions and cannot be consulted simultaneously. The stability of $K_q(\mu)$ against the running effects is also illustrated in Fig. \ref{k}.

\section{Conclusions and Discussions} \label{sec:dis}

The standard model of particle physics has achieved triumphant successes in the last five decades. However, one of the most crucial shortcomings therein is a large number of free parameters, including twelve fermion masses. Any reduction of this number will pave a way for our comprehension of the underlying flavor physics. The Koide formula is one of the appealing attempts in this direction. Unfortunately, this formula only associates the masses of three charged leptons, but not of all the twelve flavors of fermions. However, charged leptons should not be particular in fermions, so the idea to extend the original Koide formula, including all leptons and quarks on an equal footing, is thus very natural and desirable. Kartavtsev's extensions \cite{K} in Eqs. (\ref{l}) and (\ref{kq}) treated six leptons and quarks in a totally democratic manner, with a maximal $S(6)$ permutation symmetry, and a preliminary estimate indicated a certain plausibility of these extensions.

In the present paper, we explore Kartavtsev's extended Koide formulae for both leptons and quarks at length. For the lepton sector, it proves that $K_l$ cannot be equal to $2/3$ exactly with the current experimental data of the charged lepton masses and the mass-squared differences of neutrinos within $1\sigma$ confidence level. Then, our strategy is to assume the rigorous validity of the extended Koide formula for leptons and relax the uncertainty of the most inaccurate tauon mass $m_\tau$. By this means, the neutrino masses can be extracted from the extended Koide formula, if the uncertainty of $m_\tau$ is relaxed to about $2\sigma$ confidence level in both the normal and inverted mass hierarchies. The central values for three neutrino masses read: $m_1=2.06\times 10^{-4}$ eV, $m_2=8.68\times 10^{-3}$ eV, $m_3=5.02\times 10^{-2}$ eV (normal hierarchy), and $m_3=2.00\times 10^{-4}$ eV, $m_1=4.94\times 10^{-2}$ eV, $m_2=5.02\times 10^{-2}$ eV (inverted hierarchy). These results are consistent with the most stringent upper bound on the sum of neutrino masses from the measurements of the cosmic microwave background temperature spectra from the WMAP and Planck satellite experiments: $m_1+m_2+m_3<0.66$ eV ($95\%$ confidence level) \cite{planck}, and also from the data combined with the baryon acoustic oscillations: $m_1+m_2+m_3<0.23$ eV ($95\%$ confidence level) \cite{planck}. It is interesting to note that, even if the neutrino masses increase near this cosmological bound (e.g. $m_1+m_2+m_3\approx 0.23$ eV), with the constraints in Eqs. (\ref{12}) and (\ref{23}), the discrepancies between $K_l$ and $2/3$ almost remain unchanged:  $2.0\times 10^{-5}<|K_l-2/3|<3.8\times 10^{-5}$ for the normal mass hierarchy and $2.0\times 10^{-5}<|K_l-2/3|<3.9\times 10^{-5}$ for the inverted mass hierarchy. This is understandable, as the uncertainty of $K_l$ mainly comes from the uncertainties of charged leptons.

For the quark sector, the various definitions of quark masses greatly complicate the situation. The pole masses in the Koide formula become ill-defined for the light quarks, due to the non-perturbative effects in QCD at low energies. Therefore, the exploration of the extended Koide formula should only be implemented for the running quark masses. We find that the running parameter $K_q(\mu)$ is almost stable in a very large range of energy scales from GeV to $10^{12}$ GeV, mainly as a result of the large mass hierarchy in the quark sector. However, $K_q(\mu)$ is always slightly larger than $2/3$, meaning the invalidity of the extended Koide formula for the running quark masses, but this deviation is merely about $5\%$. We omit the discussion of the running behavior of the extended Koide formula in the lepton sector, as the running effects are negligibly tiny because the lepton mass ratios are rather insensitive to radiative corrections \cite{Xing:2006vk}.

\vskip .3cm

Below, we give some general discussions on the Koide formula. The mystery of the Koide formula is two-fold. The first is its surprising simplicity and accuracy, but only for charged leptons. The inclusion of neutrinos is a reasonable balance between the charged and uncharged leptons, but this inclusion is meaningful only if the generation mechanism of neutrino masses is the same as that of charged lepton masses (i.e. neutrinos are of Dirac type). On the contrary, if the tiny neutrino masses are generated from the seesaw mechanism \cite{seesaw}, Kartavtsev's extension of the Koide formula will be pointless due to the Majorana mass term. The second is that the Koide formula consists of the pole masses of fermions, which are the low energy quantities and are defined at different energy scales. This is extremely counterintuitive, since we always expect simple formulae at high energy scales, where some symmetries are restored. Hence, the renormalization effects will not allow the Koide-like formulae for both the pole and running fermion masses simultaneously.

Finally, we should point out that Kartavtsev's extension of the Koide formula \cite{K} and our corresponding detailed explorations are still at the phenomenological level. A similar work (the most general extension of the Koide formula), taking all the twelve fermions into account, i.e. $(\sum_f m_f)/(\sum_f\sqrt{m_f})^2=2/3$, is also not quite successful. Therefore, a possible direction for further extensions of the Koide-like formulae is to seek the theoretical basis of these empirical relations, as Koide originally did in a composite model or an extended technicolor-like model \cite{Koide}. We should incorporate in the extended Koide formulae the elements and the mixing and phase angles in the lepton and quark mixing matrices \cite{PMNS,CKM}, and maybe also the fermion charges. This will be the topic for our future research.

\vskip .3cm

We are very grateful to Prof. Jean-Marc G\'{e}rard for his stimulating idea and also to Fengjiao Chen for fruitful discussions. This work is supported by the Fundamental Research Funds for the Central Universities of China (No. N140504008).


\begin{thebibliography}{99}
\bibitem{Koide}
Y. Koide, Lett. Nuovo Cimento {\bf 34}, 201 (1982); Phys. Rev. D {\bf 28}, 252 (1983); Phys. Lett. B {\bf 120}, 161 (1983).

\bibitem{PDG}
K.A. Olive {\it et al.} (Particle Data Group), Chin. Phys. C {\bf 38}, 090001 (2014).

\bibitem{Gerard:2005ad}
J.-M. G\'{e}rard, F. Goffinet, and M. Herquet, Phys.\ Lett.\ B {\bf 633}, 563 (2006).

\bibitem{Koide:1989jq}
Y. Koide, Mod.\ Phys.\ Lett.\ A {\bf 5}, 2319 (1990); arXiv:hep-ph/0506247; J.\ Phys.\ G {\bf 34}, 1653 (2007); Eur.\ Phys.\ J.\ C {\bf 50}, 809 (2007); Eur.\ Phys.\ J.\ C {\bf 52}, 617 (2007); Int.\ J.\ Mod.\ Phys.\ E {\bf 16}, 1417 (2007); Phys.\ Rev.\ D {\bf 79}, 033009 (2009), Phys.\ Lett.\ B {\bf 681}, 68 (2009); Phys.\ Lett.\ B {\bf 687}, 219 (2010); Phys.\ Rev.\ D {\bf 81}, 097901 (2010).

\bibitem{Koide:1995xk}
Y. Koide and M. Tanimoto, Z.\ Phys.\ C {\bf 72}, 333 (1996);
Y. Koide and H. Fusaoka, Prog.\ Theor.\ Phys.\ {\bf 97}, 459 (1997).

\bibitem{Xing:2006vk}
Z.-z. Xing and H. Zhang, Phys.\ Lett.\ B {\bf 635}, 107 (2006).

\bibitem{Ma:2006ht}
E. Ma, Phys.\ Lett.\ B {\bf 649}, 287 (2007).

\bibitem{Sumino}
Y. Sumino, J. High Energy Phys. {\bf 0905}, 075 (2009); Phys.\ Lett.\ B {\bf 671}, 477 (2009); arXiv:0903.3640 [hep-ph].

\bibitem{Uekusa:2009pj}
N. Uekusa, Eur.\ Phys.\ J.\ C {\bf 71}, 1664 (2011).

\bibitem{z}
P. \.{Z}enczykowski, Phys.\ Rev.\ D {\bf 86}, 117303 (2012); Phys.\ Rev.\ D {\bf 87}, 077302 (2013).

\bibitem{Foot}
R. Foot, hep-ph/9402242;
A. Rivero and A. Gsponer, arXiv:hep-ph/0505220;
W. Krolikowski, arXiv:hep-ph/0508039;
N. Li and B.-Q. Ma, Phys.\ Rev.\ D {\bf 73}, 013009 (2006);
A. Rivero, arXiv:1111.7232 [hep-ph];
J. Kocik, arXiv:1201.2067 [physics.gen-ph];
F.G. Cao, Phys. Rev. D {\bf 85}, 113003 (2012).

\bibitem{Li:2005}
N. Li and B.-Q. Ma, Phys.\ Lett.\ B {\bf 609}, 309 (2005).

\bibitem{Rodejohann:2011jj}
W. Rodejohann and H. Zhang, Phys.\ Lett.\ B {\bf 698}, 152 (2011).

\bibitem{K}
A. Kartavtsev, arXiv:1111.0480 [hep-ph].

\bibitem{Koideq}
H. Fusaoka and Y. Koide, Phys.\ Rev.\ D {\bf 57}, 3986 (1998).

\bibitem{Xing}
Z.-z. Xing, H. Zhang, and S. Zhou, Phys.\ Rev.\ D {\bf 77}, 113016 (2008).

\bibitem{Xinga}
Z.-z. Xing, H. Zhang, and S. Zhou, Phys.\ Rev.\ D {\bf 86}, 013013 (2012).

\bibitem{planck}
P.A.R. Ade {\it et al.} [Planck Collaboration], Astron.\ Astrophys.\ {\bf 571}, A16 (2014).

\bibitem{seesaw}
T. Yanagida, in {\it Proceedings of the Workshop on the Unified Theory and the Baryon Number of the Universe}, edited by O. Sawada and A. Sugamoto (KEK, Tsukuba, 1979);
M. Gell-Mann, P. Ramond, and R. Slansky, in {\it Supergravity}, edited by F. van Nieuwenhuizen and D. Freedman (North Holland, Armsterdam, 1979);
R.N. Mohapatra and G. Senjanovi\'{c}, Phys. Rev. Lett. {\bf 44}, 912 (1980).

\bibitem{PMNS}
B. Pontecorvo, Sov. Phys. JETP {\bf 6}, 429 (1958); Sov. Phys. JETP {\bf 7}, 172 (1958);
Z. Maki, M. Nakagawa, and S. Sakata, Prog. Theor. Phys. {\bf 28}, 870 (1962).

\bibitem{CKM}
N. Cabibbo, Phys. Rev. Lett. {\bf 10}, 531 (1963);
M. Kobayashi and T. Maskawa, Prog. Theor. Phys. {\bf 49}, 652 (1973).


\end{thebibliography}
\end{document}